# SkyDe: a Skype-based Steganographic Method

Wojciech Mazurczyk
Warsaw University of Technology,
Institute of Telecommunications
Warsaw, Poland, 00-665,
Nowowiejska 15/19
wmazurczyk@tele.pw.edu.pl

Maciej Karaś
Warsaw University of Technology,
Institute of Telecommunications
Warsaw, Poland, 00-665,
Nowowiejska 15/19
karas.maciek@gmail.com

Krzysztof Szczypiorski
Warsaw University of Technology,
Institute of Telecommunications
Warsaw, Poland, 00-665,
Nowowiejska 15/19
ksz@tele.pw.edu.pl

## ABSTRACT
This paper introduces SkyDe (Skype Hide), a new steganographic method that utilizes Skype encrypted packets with silence to provide the means for clandestine communication. It is possible to reuse packets that do not carry voice signals for steganographic purposes because Skype does not use any silence suppression mechanism. The method's proof-of-concept implementation and first experimental results are presented. They prove that the method is feasible and offers steganographic bandwidth as high as 2.8 kbps.

## Categories and Subject Descriptors
C.2.0 [**Computer-Communication Networks**]: General, Security and protection

## 1. INTRODUCTION
Voice over IP (VoIP) or IP telephony is one of the services of the IP world that is changing the entire telecommunications landscape. It is a real-time service, which enables users to make phone calls through data networks that use an IP protocol. Currently, one of the most popular of the IP telephony systems that have shaken the global telephony market in the last decade is Skype (www. skype.com). It is a proprietary P2P telephony service originally introduced in 2003 by creators of the famous P2P file sharing system Kazaa – Niklas Zennström and Janus Friis. Skype is owned by Microsoft and it has been reported that it has about 663 million registered users (September 2011). In March 2012, it was reported to have had 35 million users online simultaneously [5]. It has also been estimated that in 2011, Skype had acquired about 33% of the world's international telephone market [18].

Steganography encompasses various information hiding techniques, the aim of which is to embed a secret message into a carrier. Network steganography, to perform hidden communication, utilizes network protocols and/or their relationships as the carrier for secret data. Because of its popularity, IP telephony is becoming a natural target for network steganography [6]. Steganographic methods are intended to hide the very existence of the communication and therefore, any third-party observers should remain unaware of the presence of the steganographic exchange.

Generally, every network steganographic method can be described by the following set of characteristics: its steganographic bandwidth, its undetectability, and robustness [1]. The term "steganographic bandwidth" refers to the amount of secret data that can be sent per unit time when using a particular method. Undetectability is defined as the inability to detect a steganogram within a certain carrier. The most popular method by which to detect a steganogram is to analyze the statistical properties of the captured data and compare them with values typical for that carrier. The final characteristic is robustness, which is defined as the amount of alteration that a steganogram can withstand without its secret data being destroyed. A good steganographic method should be as robust and as difficult to detect as possible whilst offering the highest bandwidth. However, it must be noted that there is always a fundamental trade-off necessary among these three measures.

Additionally, it is also useful to measure the steganographic cost. This is a characteristic that belongs to the sphere of carrier fidelity and has a direct impact on undetectability. It describes the degradation or distortion of the carrier caused by the application of the steganographic method. For example, in the case of VoIP steganography methods, this cost can be expressed as a measure of the conversation quality degradation induced by applying a particular technique for hiding information.

One of the most important aspects for every steganographic method is the selection of the most suitable carrier for the secret data. The most favorable carrier for secret messages must have two features:

- It should be popular, i.e., usage of such a carrier should not be considered as an anomaly in itself. The more such carriers are present and utilized on the network the better, because they mask the existence of hidden communication.

- Modification of the carrier related to the embedding of the steganogram should not be "visible" to the third party, who is unaware of the steganographic procedure.

Thus, because of its popularity and traffic volume, Skype traffic is an ideal candidate for a secret data carrier. It should also be emphasized that the purpose of establishing any information hiding exchange varies – possible uses can fall into the category of legal actions (e.g., circumvention of web censorship and surveillance, computer forensics or copyright protection) or illicit activity (e.g., criminal communication, confidential data exfiltration or industrial espionage). This trade-off is typical in steganography and requires consideration in a broader steganography context, which is beyond the scope of this paper.

One can always question why use steganography for Skype, which as is commonly known, uses cryptographic means to provide confidentiality for every type of users' message that is exchanged (text messages, voice signal, files). First of all, hidden communication need not necessarily be conducted in an end-to-end manner, i.e., covert data can be sent using third party VoIP calls. Secondly, discussion has recently arisen on whether Skype, whose calls were commonly believed very hard to wiretap, is providing lawful interception services to law enforcement agencies [7]. Additionally, Skype is proprietary and closed software and thus, ultimately cannot be trusted.

This is why in this paper we discuss how Skype can be utilized to provide means for clandestine communication. Experiments performed on real Skype traffic prove that the proposed method – SkyDe (Skype Hide) – is feasible and offers high steganographic bandwidth and reasonable steganographic cost in terms of undetectability.

The rest of this paper is structured as follows. Section 2 presents the current state of research efforts on VoIP steganography and on Skype in particular. In Section 3, Skype basics are presented, as well as an analysis of its traffic and a detailed description of the proposed method and the prototype implementation. Next, in Section 4, experimental results for the method are presented. Finally, the last section concludes our work and indicates directions for further research.

## 2. RELATED WORK

A number of steganographic approaches for VoIP have been proposed, which have been recently exhaustively surveyed by Mazurczyk [1]. Because of space limitations only steganographic methods related to Skype and silence suppression mechanisms will be mentioned here.

The only Skype-dedicated method was introduced by Wang et al. [4]. The authors proposed embedding a 24-bit watermark into the encrypted stream (e.g., Skype call) to track its propagation through the network and thus, providing its de-anonymization. The watermark is inserted by modifying the inter-packet delay for selected packets in the VoIP stream. The authors demonstrated that depending on the watermark parameters chosen, they were able to achieve a 99% true positive and a 0% false positive rate, while maintaining good robustness and undetectability. However, they achieved steganographic bandwidth of only about 0.3 bit/s, which is enough for the described application, but rather low for performing clandestine communication.

The deployment of speech codecs' Silence Insertion Description (SID) frames that are sent when voice is not present in the input signal, was proposed as a secret data carrier by Mazurczyk and Szczypiorski [2]. However, the authors noted that the steganographic bandwidth for this method is rather low, but that it is possible to influence the rate at which SID frames are issued.

Recently, a high-capacity steganography technique based on the utilization of the inactive frames of the G.723.1 speech codec was introduced by Huang et al. [3]. The authors proved that the inactive frames of VoIP streams are more suitable for data embedding than the active ones and thus, a greater amount of hidden data can be embedded within them with the same imperceptibility. They then proposed a steganographic algorithm in different speech parameters of the inactive frames for the G.723.1 codec with a 6.3 kbits bitrate. Experimental results show the solution is imperceptible and that a quite high steganographic bandwidth of up to 101 bits/frame is achieved.

Proposed in this paper, the SkyDe method operates by reusing encrypted packets with silence, by substituting this information with secret data. When compared with existing solutions, it is characterized by significantly higher steganographic bandwidth and good undetectability.

## 3. SKYPE TRAFFIC ANALYSIS AND PROPOSED METHOD DESCRIPTION
### 3.1 Skype basics

Skype is a hierarchical P2P network with a single centralized element (login server) that is formed by two types of nodes [11]:

- Ordinary Nodes (ONs) that can start and receive a call, send instantaneous messages, and transfer files, and

- Specialized nodes called Super Nodes (SNs) that are responsible for helping ONs find and connect to each other within the Skype network

The login server is responsible for the authentication of ONs and SNs before they access the Skype network.

Skype offers two communication modes, namely:, End-to-End (E2E) and End-to-Out (E2O). The first is possible between two Skype clients within the IP network. The second occurs if one of the endpoints is a Skype client within the IP network and the other is a PSTN phone (SkypeIn/SkypeOut services). In this paper, we focus solely on the Skype E2E mode.

Skype is based on proprietary protocols and makes extensive use of cryptography, obfuscation, and anti-reverse-engineering procedures, and all information about its traffic characteristics, protocols, and behaviors comes from numerous measurement studies, e.g., [8] [9].

All traffic in Skype is encrypted; neither signaling messages, nor the packets that carry voice data can be uncovered. Typically, the preferred, first-choice transport protocol of Skype is User Datagram Protocol (UDP), which as the traffic analysis in [9] showed, is being used in about 70% of all calls. However, if Skype is unable to connect using UDP it falls back to TCP. In this paper, we focus solely on UDP-based Skype calls.

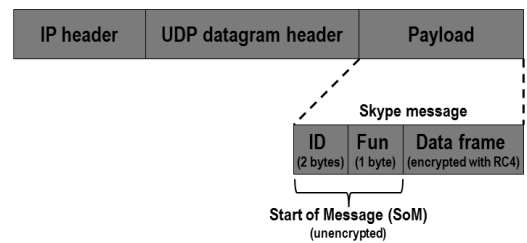

**Figure 1. UDP-based Skype message format.**

For TCP-based transport, the entire Skype message is encrypted. In cases of unreliable UDP, at the beginning of each datagram's payload, an unencrypted header is present that is called the Start of Message (SoM). It is unencrypted in order to be able to restore the sequence of packets that was originally transmitted, to detect a loss, and to quickly distinguish the type of data that is carried

inside the message. SoM consists of the following two fields [8] (Fig. 1):

- **ID** (2 bytes) that is used to uniquely identify the message; it is randomly selected by the sender query, and copied in the receiver reply.
- **Fun** (1 byte) that describes the payload type. For example, values: 0x02, 0x03, 0x07, and 0x0f are typically used to indicate signaling messages (used during the login phase or for connection management). 0x0d indicates a DATA message that can contain: encoded voice or video blocks, chat messages or fragments of files.

Skype estimates the available bandwidth and the packet loss probability, and it dynamically adapts to the detected network conditions by adjusting the codec's bitrate or by introducing higher redundancy in packets [9]. Typically, the resulting packet rate is about 16, 33, or 50 packet/s [11].

From the perspective of the steganographic method proposed in this paper, it is also important to emphasize that Skype does not utilize any silence suppression mechanism, i.e., even if there is no voice activity during the conversation, the packets that carry the silence are still generated and sent. The lack of support for silence suppression is intentional – it helps to obtain better voice quality and maintain UDP bindings at the NAT (Network Address Translation) [10].

## 3.2 Skype traffic analysis

For the purpose of the proposed steganographic method, outlined in detail in Section 3.3, an analysis of Skype traffic was carried out to prove that the method is feasible.

An experimental test-bed was set up (Fig. 2) that included two Skype clients (Skype for Linux v. 2.2.0.35 and for Windows v. 6.0.60.126) and a Linux-based application designed and developed by authors that could intercept Skype packets before they reached (for the transmitting side) and after they entered the (for the receiving side) network interface. Two instances of this application were synchronized using NTP (Network Time Protocol) and were responsible for the generation of reports regarding the Skype packets' statistics.

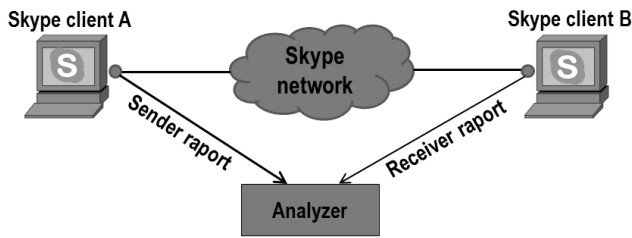

**Figure 2. Experimental Skype test-bed.**

Then, reports from both sides: transmitting and receiving, were analyzed and joined in an analyzer. This helps to observe the same Skype encrypted stream in two network localizations.

Using this test-bed, a number of measurements were carried out on Skype traffic. In each experiment, the number of analyzed Skype packets was larger than 100 000 and was repeated four times. Here, only the averaged results are presented.

### 3.2.1 Skype packets' size distribution

In 2008, Chang et al. [10] observed that in Skype traffic, speech activity is highly correlated to packet size, as more information is encoded into a voice packet while a user is speaking. Experimental results revealed that the packet size and speech volume are highly correlated because they fluctuate in tandem.

This feature of Skype traffic is utilized in SkyDe to identify the packets that carry silence. That is why we wanted to verify whether the same behavior as in [10] can be observed for current Skype software. Fig. 3 presents the obtained experimental results.

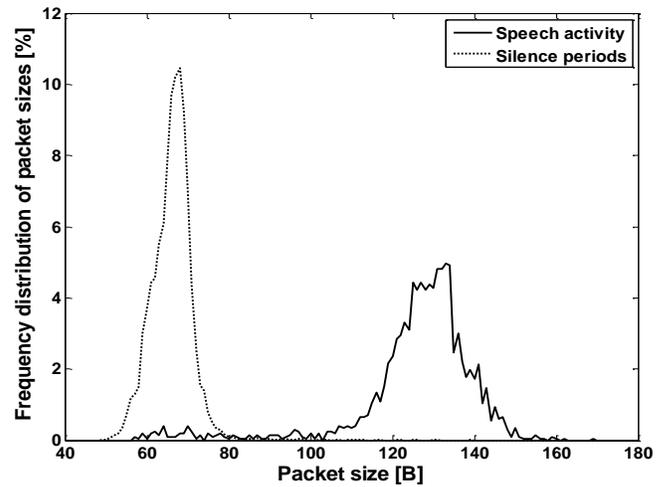

**Figure 3. The distribution of packets' size during conversation and periods of silence.**

As can be observed, the previously reported high correlation between speech activity and packet size is still evident. This means that based on the size of each Skype packet we can assess with high probability whether it carries voice or silence. As mentioned before, this effect will be also utilized by SkyDe.

### 3.2.2 Skype bytes distribution in packets' payload

To later prove SkyDe undetectability, the following characteristic of Skype bytes distribution for typical Skype calls was carried out (Fig. 4).

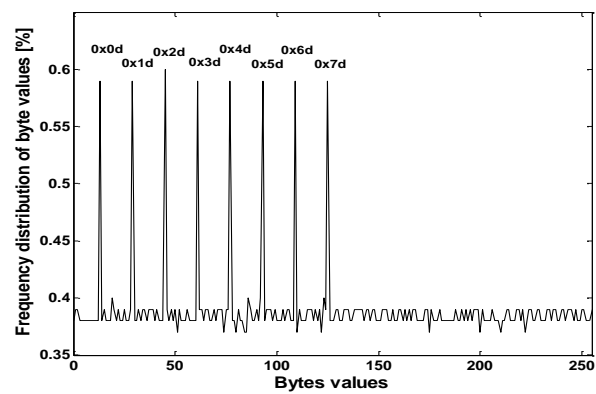

**Figure 4. The bytes distribution in packets' payload for typical Skype calls.**

It can be observed that there are certain bytes values (in hex), i.e., 0x0d, 0x1d, 0x2d, 0x3d, 0x4d, 0x5d, 0x6d, and 0x7d that are more frequent than others, whereas the rest are almost equally distributed. The popular eight bytes are the typical values from the Fun field for each UDP-based Skype packet (see Sec. 3.1). The distribution of the rest of the byte values is almost equal because they represent the encrypted payload of Skype packets.

### 3.2.3 Skype resistance to packet losses

We also investigated Skype's maximum threshold of packet losses that could be sustained by a UDP-based connection without falling back to TCP. It transpired that this value is about 70%. Of course, when this level of losses is reached, the voice quality is degenerated to the point where no further conversation is possible.

## 3.3 Proposed method and its implementation

SkyDe is designed to utilize Skype encrypted voice streams to enable clandestine communication. This secret data exchange can be realized between: (a) two Skype users, i.e., they use their own call for steganographic purposes, or (b) a steganogram transmitter and receiver that utilize an existing, third party Skype call (in this case the original caller and callee are not aware of the information hiding procedure).

SkyDe utilizes encrypted Skype voice packets as a hidden data carrier. By taking advantage of the high correlation between speech activity and packet size (described and verified in Sec. 3.2.1), packets without voice signal can be identified and used to carry secret data (by replacing the encrypted silence with secret data bits). As is commonly known, typical VoIP calls contain 35% to 70% silent periods in each direction [12], [13]. Skype does not utilize any silence suppression algorithm; thus, generally, all packets with silence can be utilized for steganographic purposes. However, SkyDe should have potentially low impact on voice quality because it does not affect packets with voice signals, which are more significant from the point of view of conversation.

### 3.3.1 SkyDe transmitting side

First, to create the secret data to be sent, as with the replaced (encrypted) data, they are encrypted prior sending. The cryptographic key utilized for encryption is a shared secret between SkyDe communication sides. Then a CRC-16 checksum is calculated on the payload and is inserted into the ID field (16 bits) of SoM (see Section 3.1). Such an approach provides steganogram integrity verification and facilitates the identification of packets that carry secret data at the receiving end.

Then, those packets that contain silence must be identified. Their size can change while the connection lasts (due to network conditions), so we propose a "sliding time window" algorithm that calculates the reference size of the packet with silence. It is also responsible for keeping the total packet losses at a safe level, because as mentioned earlier, steganographic utilization of each packet with silence will increase the overall packet loss for the call. Thus, it is important not to exceed the measured threshold of about 70% total packet losses (see subsection 3.2.3) because then the connection will fall back to TCP.

The proposed algorithm works as follows. First, the size of the "sliding time window" $w$ (in seconds) is selected, in which the reference value $r$ of Skype packets with silence is continuously updated during the call. Every second, the packet with the lowest size is determined and stored. When all of the values in the window are measured, then the average reference value is calculated based on the three packets with the lowest size. Additionally, a certain deviation in packet size ($\Delta$) $r \pm \Delta$ is acceptable for consideration as a packet for SkyDe purposes. In the same time window $w$, it is also verified whether the total packet loss level reached 70% and if it does; the steganographic method ceases to utilize packets with silence until the losses are again at a safe level.

After the packets with silence are identified, their payloads are replaced with encrypted secret data and they are sent to the receiving side.

To provide reliability for the proposed method, an additional protocol in a hidden channel might be required. One solution is to use an approach proposed by Hamdaqa and Tahvildari [17] because it can be easily incorporated into SkyDe. It provides a reliability and fault-tolerance mechanism based on a modified (k, n) threshold of a Lagrange Interpolation and the results demonstrated in that paper prove that the complexity of steganalysis is increased. Of course, the "cost" for the extra reliability would always be a loss of some fraction of the steganographic bandwidth.

### 3.3.2 SkyDe receiving side

At the receiver side, each packet with silence is recognized by the same means (packet size) as at the transmitter side. These packets are then copied to the buffer and for each of them, a CRC-16 checksum is verified and the secret data is extracted. It is not important to erase or replace secret data embedded into packets because the Skype client will treat these packets as losses. However, only the receiving end will be aware of the loss; thus, these modified SkyDe packets will not be discarded while traveling through the network.

SkyDe was implemented in the same network environment as depicted in Fig. 2, by enhancing the functionality of the application that was used to gather statistics about Skype traffic (see Section 3.2). Based on the experimental results, it transpired that SkyDe performs best when $w = 10$ seconds and $\Delta = 20$ bytes. The selection of the most optimal values of $w$ and $\Delta$ will be not covered in this paper due to space limitations.

## 4. EXPERIMENTAL RESULTS
## 4.1 Experiments methodology

For SkyDe evaluation, we prepared a .wav file that consists of male and female voices speaking English. The .wav file contains five minutes of imitated telephone conversation, i.e., sentences interleaved with periods of silence (similar to that designed in [13]) with a ratio of 55:45 (55% speech activity and 45% silence). In our experiments, we compiled the .wav file using audio recordings from the TIMIT [14] continuous speech corpus – one of the most widely used corpora in the speech recognition community. In our analysis, we use the subjective measure MOS (Mean Opinion Score) [16] calculated with the Perceptual Evaluation of Speech Quality (PESQ) method [15].

The .wav file was used as the source of the voice signal during the Skype call and was saved at the receiving end. Then, the original and degraded file were compared through the use of PESQ and the resultant MOS-LQO (MOS-Listening Quality, Objective) was returned to assess the SkyDe impact on voice quality

(steganographic cost). During the same experiments, the steganographic bandwidth and bytes distribution in the packets' payload (as in Fig. 4) were measured. Each experiment was repeated three times and averaged values are presented.

It is worth noting that during the experiments none of the calls was disconnected or reverted to TCP, which proves the correct design of the proposed method.

## 4.2 Steganographic bandwidth and cost

Using the methodology presented in the previous subsection, the results of steganographic bandwidth and cost were obtained. Steganographic bandwidth was calculated by analyzing and extracting the secret data from the Skype packets with silence that were utilized by SkyDe. Also, for these connections, the corresponding voice quality MOS-LLQ scores were determined.

First, the reference MOS-LLQ value without SkyDe applied was obtained. It transpired that the resulting quality is about 3.94, which is considered to be a good quality. The steganographic cost is then determined as a difference between the reference MOS-LLQ without steganography and that with SkyDe applied. The steganographic utilization of packets with silence was increased in a range from 0% to 100% and the obtained results regarding voice quality are presented in Fig. 5.

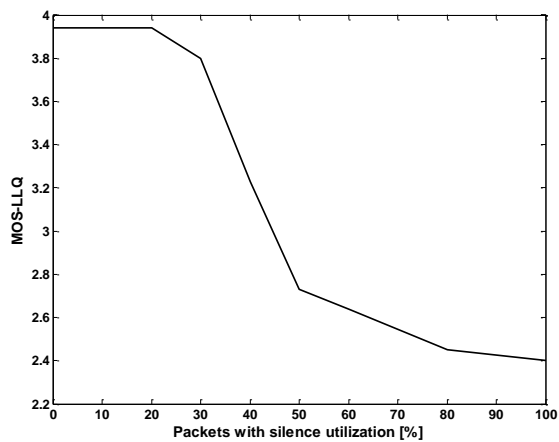

**Figure 5. Voice quality experimental results**

When up to 20% of the packets with silence are utilized for steganographic purposes, the resulting voice quality is not degraded at all. For 30%, the disruption in voice quality is negligible. For higher utilization rates, the quality begins to deteriorate more quickly and after exceeding 50% it is considered as poor; however, conversation is still possible.

**Table 1. Experimental results**

| Packets with silence utilization [%] | 0 | 20 | 30 | 40 | 50 | 60 | 80 | 100 |
|---|---|---|---|---|---|---|---|---|
| Steganographic bandwidth [kbps] | N/A | 1.37 | 1.83 | 1.52 | 1.5 | 1.81 | 2.47 | 2.78 |
| Steganographic cost | N/A | 0 | 0.14 | 0.71 | 1.21 | 1.3 | 1.49 | 1.54 |
| Packets with silence reference size [B] | 37.86 | 35.95 | 33.98 | 35.23 | 48.98 | 48.42 | 46.10 | 43.82 |
| Packet rate [packet/s] | 50.17 | 49.14 | 45.61 | 23.68 | 17.08 | 16.97 | 17.17 | 17.84 |

The results of SkyDe steganographic bandwidth and cost are presented in Table 1. With up to 30% utilization of SkyDe, the steganographic bandwidth continues to increase to finally reach about 1.8 kbps. This must be considered as a high steganographic bandwidth. For higher utilization rates, the steganographic bandwidth drops because of the decrease in packet rate – this is Skype's reaction to the elevated level of overall packet losses. For 40% utilization rate, it drops almost a half (from about 46 to 24 packet/s) and continues to decrease until it reaches levels of 17 packet/s. Simultaneously, the size of the packets increases even by 36% (for 50% to 60% silence utilization rate). This explains the drop in resulting steganographic bandwidth to the level of 1.5 kbps and then the subsequent rise to the point of about 2.8 kbits for 100% utilization rate.

Thus, the proper selection of the level of utilization of packets with silence is necessary and typically, it will be a trade-off between Skype call quality, desired steganographic bandwidth, and undetectability.

## 4.3 Undetectability

### 4.3.1 User perspective

Detection of SkyDe, as well as steganographic bandwidth, depends on the hidden communication scenario in which it is utilized. SkyDe can be used between two Skype clients that are aware of the steganographic procedure (secret data is sent in an end-to-end manner), or some intermediary nodes can rely on third party Skype calls to exchange steganograms. In the first case, users can select the desired steganographic bandwidth because they do not necessarily expect high voice quality. Additionally, they do not care about the elevated overall packet loss level; thus, they are able to transmit about 3 kbps. However, in the latter case, the degradation of voice quality or introduced losses cannot be excessive because it could make overt users suspicious. Thus, this could potentially limit the maximum steganographic bandwidth that could be achieved. Therefore, from the point of view of voice quality and packet losses, SkyDe would be most undetectable for 30% utilization of packets with silence rate (1.8 kbps of steganographic bandwidth).

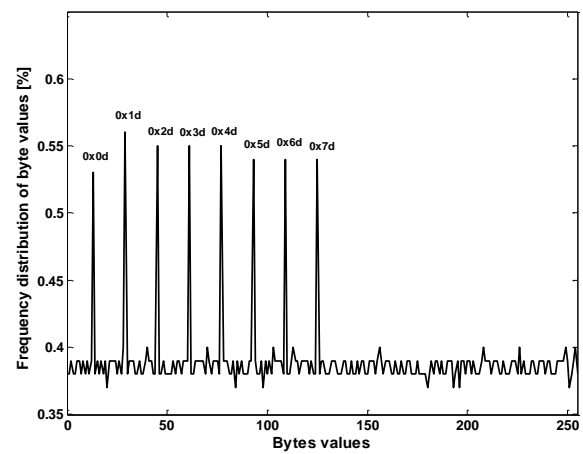

**Figure 6. The bytes distribution in packets' payload for SkyDe.**

### 4.3.2 Network perspective

To prove that the proposed method is difficult to detect, we performed similar experiments with bytes distribution in packets'

payload (Fig. 4). It was conducted for SkyDe utilizing 100% of packets with silence. The obtained results are presented in Fig. 6. To determine to what extent the results are similar to those of Fig. 4, we used correlation coefficients to quantity the strength of their relationship. It transpired that the correlation coefficient between the bytes distribution in packets' payload for typical Skype and SkyDe calls is 0.96, which indicates a very strong relationship. This makes the proposed steganographic method very difficult to detect.

However, if someone is monitoring Skype traffic, e.g., the packets rate and their sizes, then for SkyDe, higher utilization rates could be more easily visible (of course, if the monitoring entity were able to distinguish between worsening network conditions and hidden communication). Therefore, for this case also, utilization rates that mimic typical Skype call parameters should be used, i.e., up to 30% utilization of packets with silence (almost no difference in packets rate and sizes from typical Skype calls).

It should be also noted that for poorer network conditions, the more non-steganographic losses would be introduced and the less steganographic information could be transferred due to the 70% total packet losses limit.

It is also important to add that we also observed that the results could be different and fluctuate depending on the day of the week (working days or weekend), or even depending on the time of the day (office or non-office hours). These changes in Skype traffic were also discovered in other studies, e.g., in [9]. This fact also works in the favor of SkyDe because it would be difficult to establish some baseline values needed for steganalysis purposes. Therefore, this makes SkyDe more difficult to detect.

## 5. CONCLUSIONS AND FUTURE WORK

In this paper we introduced SkyDe, a steganographic method that utilizes encrypted Skype packets as a hidden data carrier. By taking advantage of the high correlation between speech activity and packet size in Skype, we identify those without voice and use them to carry secret data. Experimental results show that such an approach offers a high steganographic bandwidth of up to 1.8 kbps (for 30% utilization of packets with silence), whilst introducing almost no distortion to the Skype call. Moreover, we prove that in these circumstances, the method operates under terms of undetectability.

It should be also noted that SkyDe can be also used for other IP telephony systems that encrypt their traffic and follows the same relationship between speech activity and packet size. Chang et al. [10] point that this is also a case for UGS (Unsolicited Grant Service) [19].

Future work will include developing an improved SkyDe algorithm for the selection of packets with silence that can increase the steganographic bandwidth whilst keeping the voice distortion at a safe level. We will also conduct more in-depth studies by taking into account different parts of the week or day. We are also going to experiment with SkyDe resistance to network packet losses and their influence on the performance of the proposed steganographic method. Moreover, we will pursue the effective steganalysis method for SkyDe.

## 6. ACKNOWLEDGMENTS

This research was partially supported by the Polish Ministry of Science and Higher Education and Polish National Science Centre under grants: 0349/IP2/2011/71 and 2011/01/D/ST7/05054.